\documentclass[aps,prl,floatfix,twocolumn,reprint,amsmath,amssymb,superscriptaddress]{revtex4-1}
\usepackage{amsfonts}
\usepackage{mathrsfs}
\usepackage{amsmath}
\usepackage{color}
\usepackage{natbib}
\usepackage{graphicx}
\usepackage{bm}
\usepackage{amssymb}
\usepackage{xspace}
\usepackage{epstopdf}
\usepackage{dcolumn}
\usepackage{longtable}
\usepackage{multirow}
\usepackage[colorlinks=true, letterpaper=true, pdfstartview=FitV, linkcolor=blue, citecolor=blue, urlcolor=blue]{hyperref}

\makeatletter

\newcommand{\Rmnum}[1]{\expandafter\@slowromancap\romannumeral #1@}
\makeatother

\begin{document}
\title{Transverse Shift in Andreev Reflection}

\author{Ying Liu}
\affiliation{Research Laboratory for Quantum Materials, Singapore University of Technology and Design, Singapore 487372, Singapore}

\author{Zhi-Ming Yu}
\email{zhiming\_yu@sutd.edu.sg}
\affiliation{Research Laboratory for Quantum Materials, Singapore University of Technology and Design, Singapore 487372, Singapore}

\author{Shengyuan A. Yang}\email{shengyuan\_yang@sutd.edu.sg}
\affiliation{Research Laboratory for Quantum Materials, Singapore University of Technology and Design, Singapore 487372, Singapore}

\begin{abstract}
An incoming electron is reflected back as a hole at a normal-metal-superconductor interface, a process known as Andreev reflection. We predict that there exists a universal transverse shift in this process due to the effect of spin-orbit coupling in the normal metal. Particularly, using both the scattering approach and the argument of angular momentum conservation, we demonstrate that the shifts are  pronounced for  lightly-doped Weyl semimetals, and are opposite for incoming electrons with different chirality, generating a chirality-dependent Hall effect for the reflected holes. The predicted shift is not limited to Weyl systems, but exists for a general three-dimensional spin-orbit-coupled metal interfaced with a superconductor.
\end{abstract}
\maketitle


Spin-orbit coupling (SOC) underlies many topics that are at the frontier of current research. Intuitively, under SOC, the change of a particle's spin polarization will tend to alter its orbital motion. The effect is particularly pronounced when particles are scattered at certain interfaces. For example, when reflected at an interface, a circularly-polarized light beam acquires a transverse shift normal to its plane of incidence, known as Imbert-Fedorov shift~\cite{Fedorov1955,Imbert1972,Onoda2004,Bliokh2006,Hosten2008,Yin2013}, due to the intrinsic SOC of light~\cite{Bliokh2015}. Recently, an analogous transverse shift is predicted for Weyl electrons~\cite{Jiang2015,Yang2015b}, the fermionic cousin of photons with strong SOC in so-called Weyl semimetals~\cite{Wan2011,Murakami2007,Burkov2011,Weng2015,Huang2015,Lv2015,Xu2015,Yang2015,Xu2015a,Lv2015a}, when they are scattered by electrostatic potentials or velocity gradients. This discovery has generated great interest~\cite{Jiang2016,LWang2017}, and hints at possible universality of such effect in spin-orbit-coupled systems.

There is an intriguing scattering process unique for the normal-metal-superconductor (NS) interface---Andreev reflection~\cite{Andreev1964}, in which an incoming electron excitation from the normal metal at energy $\varepsilon$ above the Fermi level $E_F$ is reflected back as a hole excitation with energy $\varepsilon$ {below} $E_F$~\cite{Gennes1966}.
The process conserves energy and momentum but not charge: the missing charge {of} $(-2e)$ is absorbed as a Cooper pair at Fermi level into the superconductor. For excitation energies below the superconducting gap, electrons cannot penetrate into the superconudctor, and Andreev reflection becomes the dominating mechanism for transport through the NS interface. A natural question arises: Is there a transverse shift associated with Andreev reflection? This question is not trivial, since the incoming and outgoing particles possess distinct identities with opposite charges. To our knowledge, it has not been posed or studied before.

In this work, we answer the above question in the affirmative. We predict that a transverse shift generally exists in Andreev reflection between a three-dimensional spin-orbit-coupled metal and a conventional superconductor. We explicitly demonstrate the effect for two examples: a lightly-doped Weyl semimetal and a spin-orbit-coupled metal without any band-crossing. The result is derived via the quantum mechanical scattering approach, and in special cases can be exactly verified by the argument of angular momentum conservation. When symmetry argument applies, the value of the shift shows universal feature independent of the details of the scattering. For Weyl semimetals, the shifts are sizable and opposite for different chiralities, leading to a chirality-dependent Hall effect for the reflected holes. Possible experimental detection of the effect is discussed.

Let's first consider the example when the normal-metal side is a Weyl semimetal. This represents the simplest model with strong SOC, which allows a clear picture to be drawn. The essential physics learned from this example applies to more general cases. In a Weyl semimetal, an electron near a Weyl point (at $\bm K_0$) may be described by the long-wavelength model (set $\hbar=1$)
\begin{equation}\label{Weyl}
H_0=-i\chi\sum_{i=x,y,z} v_i \sigma_i\partial_i,
\end{equation}
where $\chi=\pm 1$ is the chirality of the Weyl point, $v_i$'s are the Fermi velocities, $\sigma_i$'s are the Pauli matrices corresponding to a spin or pseudospin degree of freedom, and for definiteness, we take it to be the real spin in the discussion. Weyl semimetals require breaking the product of inversion ($\mathcal{P}$) and time reversal ($\mathcal{T}$) symmetries.  We assume $\mathcal{P}$ is broken (by the underlying lattice) and $\mathcal{T}$ is preserved, then each time-reversal pair of Weyl points share the same chirality, and electrons in the $-\bm K_0$ valley are also described by Eq.(\ref{Weyl})~\cite{OtherPair}.

Consider a clean NS interface located at $z=0$ plane, with $z<0$ the Weyl semimetal (N region) and $z>0$ the superconductor (S region). The scattering at the interface is described by the Bogoliubov-de Gennes (BdG) equation~\cite{Gennes1966,Blonder1982}, in which the electron excitation at one valley is coupled to the hole excitation at the time-reversed valley by the superconducting pair potential. Focusing on the scattering of an incident electron at $\bm K_0$ valley, its BdG equation reduces to the following form when intervalley scattering can be disregarded
\begin{equation}\label{WBdG}
\left[
  \begin{array}{cc}
    H_0+U(\bm r)-E_F & \Delta(\bm r) \\
    \Delta^*(\bm r) & E_F-H_0-U(\bm r) \\
  \end{array}
\right]\psi=\varepsilon\psi.
\end{equation}
Here $\psi\equiv(\psi_{+,\uparrow},\psi_{+,\downarrow},\psi_{-,\downarrow}^*,-\psi_{-,\uparrow}^*)^T$ is the four-component spinor wave-function, the first two components represent the electron spinor in the $\bm K_0$ valley, while the latter two form the hole spinor at $-\bm K_0$ valley;  they are coupled by the pair potential on the S side. For simplicity, we model the S side using the same Weyl Hamiltonian $H_0$ but is heavily doped (represented by the potential $U$)~\cite{Meng2012}. We emphasize that this choice is only for making the problem analytically solvable, and is \emph{not necessary} for the essential physics. As we shall see, S region modeled based on the free electron model will give the same result of transverse shift. We adopt the usual step-function model for the pair potential $\Delta(\bm r)=\Delta_0e^{i\varphi}\Theta(z)$ and the single-body potential $U(\bm r)=-U_0\Theta(z)$ with $\Theta$ the Heaviside step function~\cite{Blonder1982,Jong1995}, assuming that the length-scale of potential variation is small compared with the Fermi wavelength of N region, but is still larger than the lattice scale. For a single NS interface, the superconducting phase $\varphi$ can always be gauged away. The mean-field requirement of superconductivity is that $E_F+U_0\gg \Delta_0$, i.e., the Fermi wavelength of S region should be much smaller than the coherence length.

Scattering states of Eq.~(\ref{WBdG}) can be solved in the standard way~\cite{Beenakker2006}. Assuming electron-doped case on the N side ($E_F>0$), an incident electron wave with excitation energy $\varepsilon$ can be expressed as
$
\psi^{e+}=\frac{1}{\sqrt{1-\eta_e^2}}(e^{-i\alpha/2},\eta_e e^{i\alpha/2},0,0)^T e^{ik_x x+ik_y y+i k_z^e z}
$.
The corresponding reflected electron and hole states are fixed by the conservation of energy and transverse momentum $\bm k_\|=(k_x,k_y)$, with $\psi^{e-}=\frac{1}{\sqrt{1-\eta_e^2}}(\eta_e e^{-i\alpha/2},e^{i\alpha/2},0,0)^T e^{ik_x x+ik_y y-i k_z^e z}$, and $\psi^{h-}=\frac{1}{\sqrt{1-\eta_h^2}}(0,0,e^{-i\alpha/2},\eta_h e^{i\alpha/2})^T e^{ik_x x+ik_y y+i k_z^h z}$. Here $k_z^{e/h}={\text{sgn}(E_F\pm \varepsilon)}\sqrt{(E_F\pm\varepsilon)^{2}-v_x^2 k_x^2-v_y^2 k_y^2}/{v_z}$, $\alpha=\arctan(\frac{v_yk_y}{v_x k_x})$, and $\eta_{e/h}=\chi \text{sgn} (E_F\pm\varepsilon)\sqrt{\frac{E_F\pm\varepsilon-\chi v_z k_z^{e/h}}{E_F\pm\varepsilon+ \chi v_z k_z^{e/h}}}$. We have chosen the normalization factors such that each basis state carries the same particle current along $z$. The reflected states $\psi^{e-}$ and $\psi^{h-}$ are connected to the incident state via the reflection amplitudes $r$ and $r_A$ respectively, which can be solved by matching the boundary condition at $z=0$ along with the basis states on the S side. Straightforward calculation~\cite{SI} leads to $r=-X^{-1}(\eta_{e}-e^{-2i\chi \beta}\eta_{h})$ for normal reflection, and
\begin{equation}
r_A=X^{-1}\sqrt{(1-\eta_e^2)(1-\eta_h^2)}e^{-i \chi \beta}
\end{equation}
for Andreev reflection,
where $X=1-e^{-2i\chi \beta}\eta_{e}\eta_{h}$, $\beta=\arccos(\frac{\varepsilon}{\Delta_0})$ for $\varepsilon<\Delta_0$, whereas $\beta=-i\text{arcosh}(\frac{\varepsilon}{\Delta_0})$ for $\varepsilon>\Delta_0$. One verifies that $|r|^2+|r_A|^2=1$ for $\varepsilon<\Delta_0$, as required by the quasiparticle current conservation.

\begin{figure}[t]
\includegraphics[width=8.5cm]{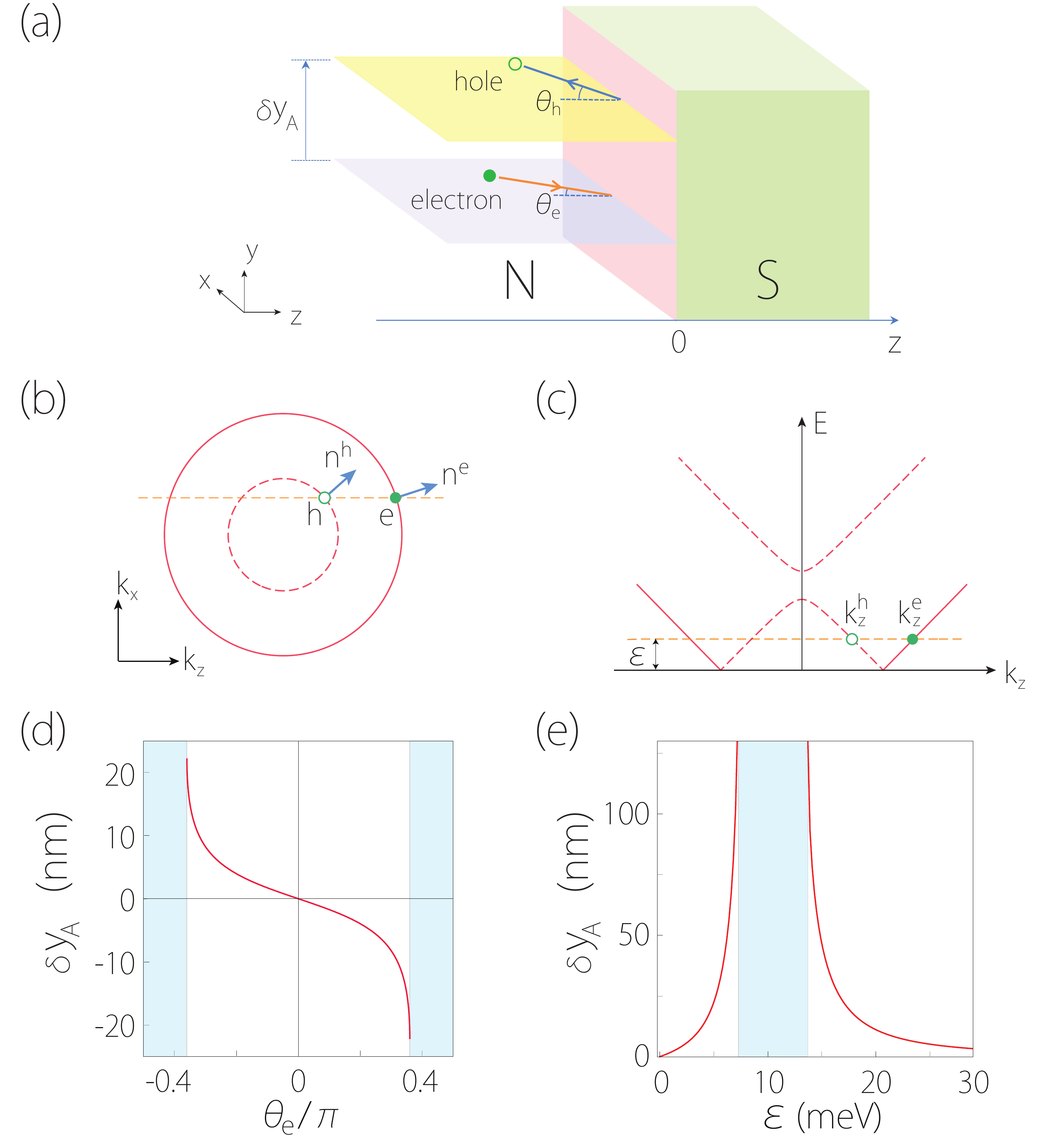}
\protect\caption{(a) Schematic figure showing the transverse shift $\delta y_{A}$ for an incident electron wave-packet in the $x$-$z$ plane Andreev reflected at the NS interface. (b,c) Schematic figure showing (b) the BdG Fermi surfaces,  and (c) spectrum at a finite $k_x$ [corresponding to the horizontal dashed line in (b)]. The solid (hollow) sphere denotes the incident electron (reflected hole) state, and the arrows indicate their spin directions. (d,e) Transverse shift versus (d) incident angle $\theta_e$ and (e) excitation energy. Here we take $\chi=+1$, $v_{x/y/z}=1.5\times 10^{6}\ \rm{m/s}$ and $E_F=10\ \rm{meV}$, $\varepsilon=E_F/20$ in (d), and $\theta_e=-\pi/20$ in (e).
} \label{Fig. 1}
\end{figure}

The spatial shift in reflection can be calculated by tracing the trajectory of a wave-packet~\cite{Xiao2010} constructed from the scattering states. Following Jiang \emph{et al.}~\cite{Jiang2015}, we obtain both longitudinal (analogous to Goos-H\"{a}nchen shift in optics~\cite{Goos1947}) and transverse shifts for normal and Andreev reflections.
The result for normal reflection recovers that in Refs.~\cite{Jiang2015,Yang2015b}. Here we focus on the shift in Andreev reflection, obtained as (including both longitudinal and transverse components)~\cite{SI}
\begin{equation}\label{ARshift}
\delta r_{i}^c=-\left[\frac{1}{2}\left(\frac{1-\eta_e^2}{1+\eta_e^2}-\frac{1-\eta_h^2}{1+\eta_h^2}\right)\frac{\partial \alpha}{\partial k_i}+
\frac{\partial \phi_A}{\partial k_i}\right]_{\bm k_\|={\bm k}_{\|}^c},
\end{equation}
where $i\in \{x,y\}$, $({\bm r}^c, {\bm k}^c)$ denotes phase-space center of the wave-packet, and $\phi_A\equiv\arg(r_A)$ is the phase of $r_A$. For simple notations, the superscript $c$ will be dropped wherever appropriate.
As sketched in Fig.~\ref{Fig. 1}(a), considering the plane of incidence to be the $x$-$z$ plane, the transverse shift $\delta y_A$ for Andreev reflection can be obtained with a simple expression
\begin{equation}\label{Wshift}
\delta y_A=\frac{\chi}{2}\frac{v_y v_z}{v_x}\left(\frac{\cot\theta_h}{E_F-\varepsilon}-\frac{\cot\theta_e}{E_F+\varepsilon}\right),
\end{equation}
where $\theta_{e/h}=\arctan(\frac{{k}_x}{k_z^{e/h}})$. Before analyzing its details, one notes a remarkable character of this result: it is independent of $\Delta$, which is actually the cause of Andreev reflection. This strongly suggests a possible symmetry interpretation.

Indeed, the total angular momentum of a system must conserve along the direction of rotational symmetry. As the transferred Cooper pair (for conventional superconductor) carries zero angular momentum, the total angular momentum $J_z$ of the quasiparticle wave-packet must be conserved during the scattering in the presence of rotational symmetry along $z$~\cite{Onoda2004,Yang2015b}. Specializing to model (\ref{WBdG}), this happens when $v_x=v_y$, and the total angular momentum of the quasiparticle wave-packet is given by~\cite{SI}
\begin{equation}
\bm J={\bm r}^c\times {\bm k}^c+\frac{\chi}{2}\bm n,
\end{equation}
where the two terms represent the orbital and spin angular momenta respectively, the unit vector $\bm n$ is the spin-polarization direction with
$\bm n^{e/h}=(v_x k_x,v_y k_y, v_z k_z^{e/h})/(E_F\pm \varepsilon)$ for electron and hole, respectively.  Note that the hole spin is opposite to that of the corresponding electron state, since a hole represents a missing electron. Thus, any change of spin polarization $\bm n$ in scattering (see Fig.~\ref{Fig. 1}(b)) necessarily requires a compensating shift in the orbital motion, in order to guarantee the $J_z$-conservation.
For the configuration in Fig.~\ref{Fig. 1}(a,b), conservation of $J_z$ immediately leads to
\begin{equation}\label{sym}
\delta y_A=\frac{\chi}{2 k_x}(n_z^h-n_z^e)=\frac{\chi}{2}v_z\left(\frac{\cot\theta_h}{E_F-\varepsilon}-\frac{\cot\theta_e}{E_F+\varepsilon}\right),
\end{equation}
exactly recovering Eq.~(\ref{Wshift}) when $v_x=v_y$~\cite{Eq5}.

The symmetry argument helps to clarify features of the shift. As shown in Fig.~\ref{Fig. 1}(d), $\delta y_A$ is an odd function of the incident angle $\theta_e$; it vanishes at normal incidence where $\bm n_e$ and $\bm n_h$ are parallel, and reaches maximum magnitude at  $\theta_e^c=\pm\arctan (\frac{v_z}{v_x}\frac{|E_F-\varepsilon|}{2\sqrt{E_F\varepsilon}})$, beyond which $k_z^h$ becomes imaginary and electrons can no longer be Andreev reflected (corresponding to the shaded regions in Fig.~\ref{Fig. 1}(d,e)). $\delta y_A$ vanishes when $\varepsilon\ll E_F$ or $\varepsilon\gg E_F$, because $\bm n_e$ and $\bm n_h$ become parallel in both limits; and its seemingly divergent behavior at $\varepsilon\rightarrow E_F$ is reconciled by noting that in this limit the hole Fermi surface becomes a point, so the reflection has a vanishingly small probability. In fact, $\varepsilon=E_F$ marks the transition point between Andreev retroflection ($\theta_e\theta_h>0$) and specular reflection ($\theta_e\theta_h<0$), as first studied in graphene~\cite{Beenakker2006,Beenakker2008,Zhang2008}. Here we find that $\delta y_A$ has the same sign in both regimes. Importantly, the shift is opposite for different chirality, i.e., the left- and right-handed holes shift in opposite transverse directions, generating a chirality-dependent Hall effect for the Andreev-reflected holes, similar to that for the normal reflection~\cite{Yang2015b} but being dominating for excitation energies below the superconducting gap~\cite{SI}.

More importantly, the symmetry argument demonstrates that the transverse shift is {independent} of the details of the NS interface and of the S region.
This confirms our previous claim regarding the modeling of S side: the Weyl-like model and SOC are \emph{not necessary} for the S region; any conventional superconductor would work. As long as the rotational symmetry is maintained (in the long-wavelength model), the same $\delta y_A$ will result from the $J_z$-conservation and only depend only on the SOC of the N side. Even if rotational symmetry is broken, a nonzero shift due to the coupled spin and orbital dynamics should generally be expected, which can be calculated using the scattering approach outlined here.

\begin{figure}[t]
\includegraphics[width=9cm]{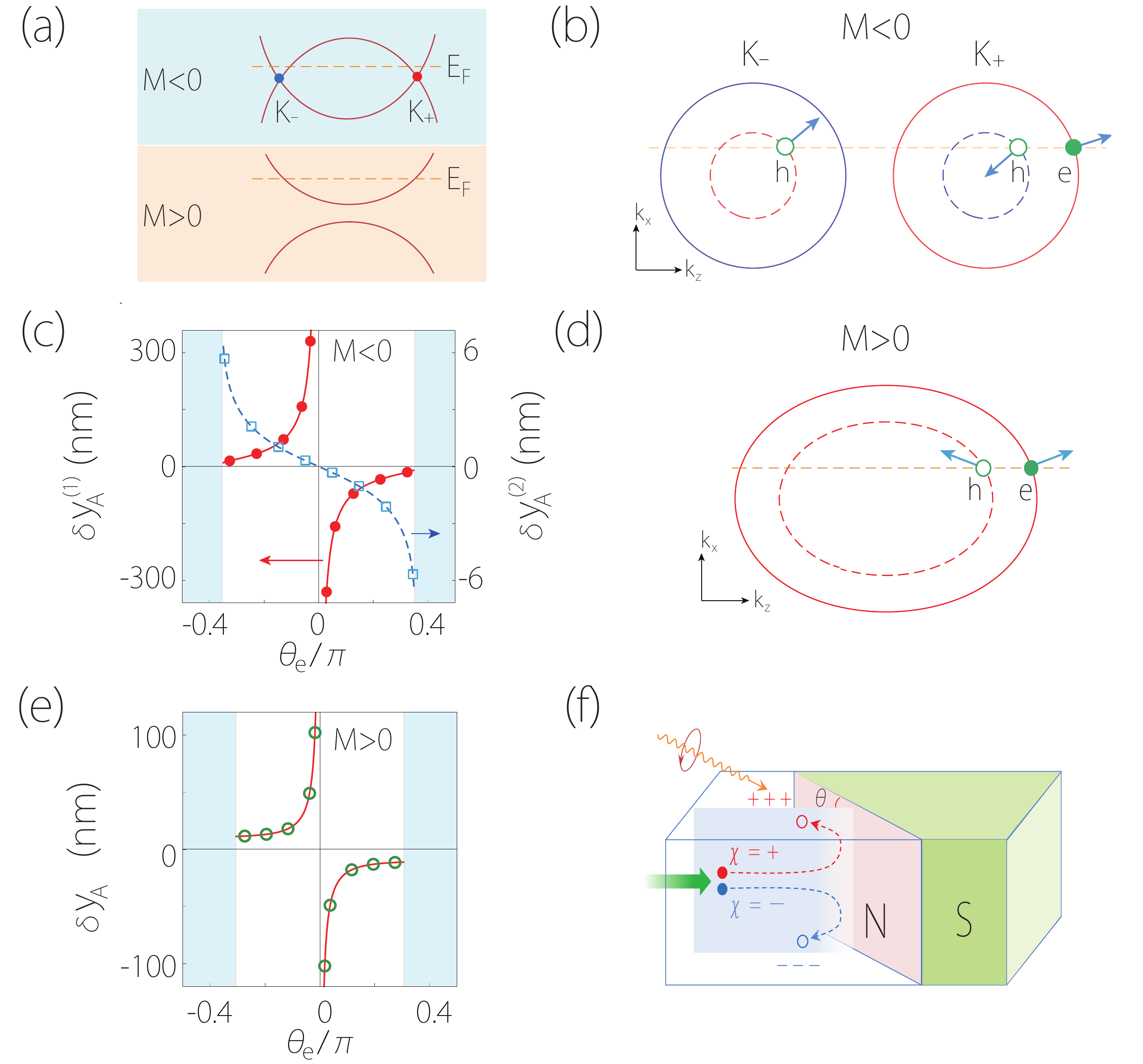}
\protect\caption{
(a) Two phases of model (\ref{M2}). Their corresponding BdG Fermi surfaces are schematically shown in (b) $M<0$, and (d) $M>0$. (c) Shift $\delta y_{A}^{(1)}$ ($\delta y_{A}^{(2)}$) for intravalley (intervalley) Andreev reflection versus the incident angle $\theta_e$ for $M<0$, with incident electron from the $K_+$ valley. (e) shows the corresponding result for $M>0$. In (c,e), the data points are from scattering approach, while the curves are from symmetry argument. Here $v=1.5\times10^{6}\ \rm{m/s}$, $m_\text{N/S}=0.04 m_e$; and $M=-0.18 m_{\text N}\cdot\text{eV}$, $E_F=20\ \rm{meV}$, $\varepsilon=0.5\ \rm{meV}$ in (c); and $M=0.18 m_{\text N}\cdot\text{eV}$, $E_F=120\ \rm{meV}$, $\varepsilon=1\ \rm{meV}$ in (e).
(f) Chirality accumulation of Andreev reflected holes on the top and bottom surfaces for a $\mathcal{T}$-preserved Weyl semimetal, which can be detected by the imbalanced absorbance of the left and right circularly polarized lights.}
\label{Fig. 2}
\end{figure}

The above-mentioned points are further illustrated with the following example. Now we take for the N side a two-band model
\begin{equation}\label{M2}
H_0=\frac{1}{2m_{\text N}}(-\nabla^2+M)\sigma_z-iv\sigma_x\partial_x-iv\sigma_y\partial_y,
\end{equation}
which nicely interpolates between two distinct phases determined by the sign of $M$ (Fig.~\ref{Fig. 2}(a)): for $M<0$, it hosts a pair of Weyl points on $k_z$-axis at $\pm\sqrt{-M}$ with \emph{opposite} chirality, simulating a $\mathcal{T}$-broken Weyl semimetal~\cite{Yang2011c,Uchida2014}; for $M>0$, the two bands are fully separated with a gap, and when $E_F> M/2m_{\text N}$, it becomes a spin-orbit-coupled metal. For the S side, we take it to be the simplest metallic superconductor \emph{without} SOC. When written in BdG equation, it appears as
\begin{equation}
\mathcal{H}_{\text S}=\left[(-\frac{1}{2m_{\text S}}\nabla^2-U_0-E_F)\tau_z+\Delta_0 \tau_x\right]\otimes \sigma_0,
\end{equation}
where $\tau_i$'s are the Pauli matrices acting on Nambu space, and $\sigma_0$ is the identity matrix in spin space. Possible interfacial barrier can also be modeled by adding a potential $h\delta(z)$ at the interface~\cite{Blonder1982}.

The transverse shift can be calculated using scattering approach like in the first example. The obtained numerical results
are shown in Fig.~\ref{Fig. 2}(c,e). First, consider the Weyl semimetal case ($M<0$). At low-energy with $|E_F+\varepsilon| \ll |M/2m_{\text N}|$, the quasiparticles are described by the Weyl model $H_\pm=-iv\sigma_x\partial_x-iv\sigma_y\partial_y\mp iv_z \sigma_z\partial_z$, where $\pm $ denotes the two valleys (also corresponding to their respective chirality $\chi$), and $v_z={\sqrt{-M}}/{m_{\text N}}$. This model allows an explicit account of intervalley scattering effect~\cite{LWang2017}: the reflected electron and hole can be either intravalley or intervalley.  Here we focus on the shifts in Andreev reflection; the shifts in normal reflection can be found in \cite{SI}.
Importantly, the intravalley hole band is of opposite chirality to the incident electron band, such that the corresponding change in spin angular momentum becomes larger (see Fig.~\ref{Fig. 2}(b)). With rotational symmetry along $z$, the shift for intravalley Andreev reflection can be obtained as
$
\delta y_A^{(1)}=-\frac{\chi}{2}v_z\left(\frac{\cot\theta_h}{E_F-\varepsilon}+\frac{\cot\theta_e}{E_F+\varepsilon}\right),
$
whereas the result $\delta y_A^{(2)}$ for intervalley Andreev reflection is the same as Eq.~(\ref{sym}) (here $\chi$ is for the incident electron). $\delta y_A^{(1)}$ and $\delta y_A^{(2)}$ have the same sign and $|\delta y_A^{(1)}|>|\delta y_A^{(2)}|$ (see Fig.~\ref{Fig. 2}(c)). These results from symmetry argument agree perfectly with the numerical results.

Next, consider the metal case for $E_F> M/2m_{\text N}>0$. On the N side, we now have a single electron Fermi surface, and the change of spin direction in Andreev reflection is illustrated in Fig.~\ref{Fig. 2}(d), which necessitates the presence of a nonzero transverse shift. We find that
$
\delta y_A=\frac{1}{2k_x}(n^h_z-n^e_z)
$, with $n^{e/h}_z=\pm[(E_F\pm \varepsilon)^2-v^2 k_\|^2]^{\frac{1}{2}}/(E_F\pm \varepsilon)$ (whereas the shift in normal reflection vanishes).
Again, this result perfectly agrees with our numerics (Fig.~\ref{Fig. 2}(e)).

These results explicitly demonstrate the following. (i) The key ingredient for the shift is SOC on the N side, however, Weyl or other types of band-crossings are not necessary. (ii) The role of the S side is to enable the electron-hole conversion. Any conventional superconductor suffices and it does not require SOC. (iii) Factors such as intervalley scattering, interfacial barrier, Fermi surface mismatch, and spatial profile of pair potential are inessential for the shift. And when symmetry argument applies, they have \emph{no} effect on the value of the shift (as defined for a definite scattering process of the wave-packet), although they do affect the probability of the process~\cite{SI,Bovenzi2017}.

A few remarks are in order.  First, distinct from the shifts in optics and in normal electron scattering, the shift discovered here occurs in a process where particle identity is completely changed, which is highly nontrivial. Unlike those usual reflections, Andreev reflection fundamentally involves two particles: the reflected hole is from the second electron that is partnered with the incident electron, and the correlation between them is established through the superconductor.
This unique character leads to features distinct from other
related shifts: the incident and outgoing particles are occupying different bands and may be tuned independently, which strongly affects the resulting shift. This is clearly illustrated in the two examples studied here, where the hole band can have chirality independent of the incident electron band. Consequently, the shift in Andreev reflection shows distinct characteristics depending on the symmetry breaking, in sharp contrast to the shift in normal reflection. 

Second, we assumed $\sigma$ as real spin in the discussion. The arguments apply equally well for pseudospins, as long as they are coupled with the orbital motion and change in the scattering. Particularly, the results for the Weyl model here directly applies for those spin-orbit-free Weyl semimetals~\cite{Weng2015a,Chen2015}.

Third, the S side is assumed to be a conventional superconductor in this work. The study can be directly extended to unconventional superconductors. One expects that interesting physics might happen when the transferred Cooper pair has finite angular momentum, which could affect the shift in Andreev reflection.

Finally, we comment on possible experimental probe of the effect. For Weyl semimetals, regardless of being $\mathcal{P}$-broken or $\mathcal{T}$-broken, the transverse shift will lead to a spatial separation of the left-handed and right-handed holes from Andreev reflection of an collimated incident electron beam, e.g., by using the geometry in Fig.~\ref{Fig. 2}(f). The excitation energy can be controlled by applied bias voltage. With preserved $\mathcal{T}$, the Andreev reflection generally dominates for excitation energies below the gap, and its probability is usually maximized when $\varepsilon\lesssim\Delta_0$~\cite{Blonder1982,SI}. The resulting surface chirality accumulation near the NS interface can be probed by the imbalanced absorbance of the left and right circularly polarized lights~\cite{Hosur2015}. In the case of a $\mathcal{T}$-broken spin-orbit-coupled metal as in model (\ref{M2}) with $M>0$, the shift generates a voltage difference between the top and bottom surfaces. Note that the bulk anomalous Hall effect does not contribute to such a voltage when the system has two-fold rotational axis along $z$ (as in model (\ref{M2})), which ensures a vanishing $\sigma_{yz}$.

\begin{acknowledgements}
\end{acknowledgements}

\bibliography{Tshift_ref}

\begin{thebibliography}{40}
\bibitem [{Oth()}]{OtherPair}%
  \BibitemOpen
  \href@noop {} {}\bibinfo {note} {There is at least another pair of Weyl
  points with chirality $-\chi$. As long as they are sufficiently separated in
  $k$-space, their presence will not affect the following
  analysis.}\BibitemShut {Stop}%
\bibitem [{Eq5()}]{Eq5}%
  \BibitemOpen
  \href@noop {} {}\bibinfo {note} {Even if $v_x\neq v_y$, Eq.~(\ref{Wshift})
  can still be reproduced by the argument combined with a scaling of
  coordinates: $(x',y')=(x\sqrt{v_y/v_x},y\sqrt{v_x/v_y})$.}\BibitemShut
  {Stop}%
\end{thebibliography}

\end{document}